\documentclass[twocolumn,amssymb,aps,floatfix]{revtex4-2}
\emergencystretch=\maxdimen
\usepackage{graphicx}
\usepackage{dcolumn}
\usepackage{amsmath}
\usepackage[latin1]{inputenc}
\usepackage{amssymb}
\usepackage[colorlinks=true, citecolor=blue, urlcolor = blue, linkcolor= red, bookmarks=true]{hyperref}
\usepackage{float}
\usepackage{tabularray}
\usepackage{amsfonts}
\usepackage{dcolumn}
\usepackage{hyperref}
\usepackage{subfigure}
\usepackage{pgfplots}
\usepackage{booktabs}
\usepackage{mathrsfs}
\usepackage{multirow}
\usepackage{afterpage}
\pgfplotsset{compat=1.16}

\usepackage{wrapfig}
\usepackage{adjustbox}
\makeatletter
\def\btt#1{\texttt{\@backslashchar#1}}
\DeclareRobustCommand\bblash{\btt{\@backslashchar}} \makeatother
\begin{document}
\title[]{Strong Gravitational Lensing by Rotating Quantum-Corrected Black Holes: Insights and Constraints from EHT Observations of M87* and Sgr A*}

\author{Amnish Vachher}\email{amnishvachher22@gmail.com} 
\affiliation{Centre for Theoretical Physics, 
	Jamia Millia Islamia, New Delhi 110025, India}
\author{Sushant~G.~Ghosh }\email{sghosh2@jmi.ac.in}
\affiliation{Centre for Theoretical Physics, 
	Jamia Millia Islamia, New Delhi 110025, India}
\affiliation{Astrophysics and Cosmology Research Unit, 
	School of Mathematics, Statistics and Computer Science, 
	University of KwaZulu-Natal, Private Bag 54001, Durban 4000, South Africa}
 
\begin{abstract}
\begin{center}
{\bf Abstract}
\end{center}
We study gravitational lensing in the strong-field limit using the rotating quantum-corrected black hole (RQCBH) with an additional parameter $\alpha$ besides mass $M$ and spin parameter $a$. We discover a decrease in the deflection angle $\alpha_D$, the photon sphere radius $x_{ps}$, and the angular position $\theta_{\infty}$. The flux ratio of the first image to all subsequent images, $r_{mag}$, decreases rapidly as $\alpha$ increases. We compare RQCBH observables with those of Kerr black holes, using Sgr A* and M87* as lenses to observe the effect of the quantum-corrected parameter $\alpha$. For Sgr A*, the angular position $\theta_\infty$ in $\in~(14.8-26.3)~\mu as$, while for M87* $\in~(11.12-19.78)~\mu as$. The angular separation $s$, for supermassive black holes (SMBHs) SgrA* and M87*, differs significantly, with values ranging $\in~(0.033-0.79)~\mu as$ for Sgr A* and $\in~(0.033-0.59)~\mu as$ for M87*. The deviations of the lensing observables $|\Delta\theta_\infty|$ and $|\Delta s|$ for RQCBH ($a=0.8,\alpha=0.4$) from Kerr black holes can reach up to $1.6~\mu as$ and $0.41~\mu as$ for Sgr A*, and $1.2~\mu as$ and $0.31~\mu as$ for M87*. The relative magnitude $r_{mag}$ $\in~(1.81-6.82)~\mu as$. We also compared the time delays between the relativistic images in the 22 SMBHs at the center of various galaxies. We found that RQCBH can be quantitatively distinguished from Kerr black holes. Interestingly, the time delay for Sgr A* and M87* can reach approximately  {6.0127 min} and 308.15 hrs, respectively. Our analysis concludes that, within the 1$\sigma$ region, a significant portion of the parameter space agrees with the EHT results of M87* and Sgr A*.  
\end{abstract}

\pacs{ } \maketitle
\section{Introduction}
Black holes, one of the most intriguing predictions of General Relativity (GR), serve as natural laboratories for testing gravity in its most robust regime \cite{Yagi:2016jml}. GR has passed numerous observational and experimental tests, from solar system observations to cosmological scales \cite{Will:2014kxa,Psaltis:2018xkc}. Astrophysical black holes in our Universe are predicted to be described by the Kerr metric, which represents rotating black holes in GR. However, many researchers study black hole solutions in modified theories of gravity (MTG) \cite{Berti:2015itd}, driven by the need to explain phenomena such as the accelerating expansion of the Universe, which cannot be fully explained by GR alone \cite{Clifton:2011jh,Nojiri:2010wj,Koyama:2015vza,Bozza:2002zj,telescope2022collaboration}. In addition, no direct observational evidence confirms that all black holes are definitively Kerr black holes, and current observations, including gravitational wave detections like GW150914 \cite{LIGOScientific:2016aoc,LIGOScientific:2016fpe}, leave room for the possibility that the geometry of these black holes might differ significantly from the Kerr metric.
Among the most promising astrophysical phenomena for testing black hole geometry and GR in general is gravitational lensing, which has long been recognized as a valuable tool for testing the predictions of GR \cite{Virbhadra:1999nm,Perlick:2004tq}. In recent years, significant theoretical attention has been focused on strong-field gravitational lensing, where light rays pass very close to the event horizon of a black hole, allowing for a deeper investigation of the spacetime geometry \cite{Bozza:2010xqn,Bozza:2002af}. Observing gravitational lensing around a black hole provides a powerful way to test GR and potential deviations introduced by modified gravity or quantum corrections. The recent breakthrough of capturing images of a black hole using the Event Horizon Telescope (EHT) has revealed that gravitational lensing is crucial for studying isolated, dim black holes such as M87* and Sgr A* \cite{EventHorizonTelescope:2019dse,EventHorizonTelescope:2022xqj}.   For the astrophysical phenomena of black holes, gravitational lensing in the strong-field regime provides unique opportunities to study spacetime geometry near the event horizon. The study of gravitational lensing has a long and fascinating history, beginning with the development of GR itself by Einstein \cite{Einstein:1936llh}, and it has significantly enhanced our understanding of spacetime \cite{Refsdal:1964yk, Liebes:1964zz, Mellier:1998pk, Bartelmann:1999yn,Perlick:2003vg,Perlick:2004tq, Schmidt:2008hc,Perlick:2010zh, Barnacka:2013lfa}.

Strong gravitational lensing, where large angles deflect light due to the intense gravitational field near a black hole,  Darwin \cite{darwin1959gravity} was one of the earliest to apply gravitational lensing concepts to Schwarzschild black holes and derived the strong-deflection limit for light rays moving near the photon sphere. This logarithmic approximation for the bending angle near the photon sphere laid the foundation for future developments in the field and significant contributions to the study of gravitational lensing by black holes \cite{Synge:1966okc,Cunha:2018acu,Luminet:1979nyg}. The theoretical work by Virbhadra and Ellis \cite{Virbhadra:1999nm,Virbhadra:2002ju}, focused on the formation and location of relativistic images around black holes, further improving our knowledge of gravitational lensing in strong-field regimes.
 The EHT observation images revealed the shadow of the black holes, corresponding to the photon ring- a bright boundary surrounding the central dark region where photon orbits become trapped under the influence of the strong gravitational field. The shadow size and shape are consistent with the predictions from the Kerr black hole solution, which describes rotating black holes within GR \cite{Bozza:2010xqn, Gralla:2020yvo, Gralla:2020srx, Johnson:2019ljv}. However, deviations from the Kerr solution due to quantum corrections, or alternative gravity theories, could manifest themselves in the precise structure of the black hole shadow and, more generally, in the strong gravitational lensing effects \cite{Gralla:2019drh, Tsupko:2022kwi,Kocherlakota:2024hyq}.
 
The study of gravitational lensing in modified black hole spacetimes has been an active area of research, mainly to test deviations from GR. Frittelli, Killing, and Newman \cite{Frittelli:1999yf} introduced a more precise analytical approach for analyzing the lens equation and its solutions.
 Further advancements were made by Bozza, who developed the strong-deflection limit method to analyze gravitational lensing near black holes \cite{Bozza:2001xd, Bozza:2002zj, Bozza:2002af, Bozza:2007gt}. These methods have been extended to various black hole spacetimes, including Reissner-Nordstr\"{o}m black holes \cite{Eiroa:2002mk}, spherically symmetric spacetimes \cite{Perlick:2004tq}, and rotating black holes \cite{Bozza:2002af}.    Building on Bozza's \cite{Bozza:2002af} analytical formulas, which were based on the strong field limit approximation, researchers have applied these methods to various black hole scenarios, including Reissner-Nordstr\"{o}m and braneworld black holes \cite{Eiroa:2003jf, Whisker:2004gq, Eiroa:2004gh, Eiroa:2012fb, Bhadra:2003zs}. 
  Recent research has also focused on strong gravitational lensing in alternative theories of gravity, such as higher curvature gravity \cite{Kumar:2020sag, Islam:2020xmy, Narzilloev:2021jtg,Feleppa:2024vdk} and string theories \cite{Bhadra:2003zs, Sharif:2017ogw, Younesizadeh:2022czv}. These studies have explored how modifications to the Schwarzschild and Kerr geometries can influence lensing observables such as the deflection angle, the photon sphere, and relativistic images \cite{Synge:1966okc, Cunha:2018acu, Shaikh:2019itn}. The calculations of positions and magnifications of higher-order images have already been carried out for Schwarzschild black holes \cite{Bozza:2001xd} and later extended to generic spherically symmetric spacetimes \cite{Bozza:2002zj}. Extending these methods to rotating black holes has allowed detailed predictions of strong lensing in Kerr spacetimes \cite{Bozza:2002af}. Tsukamoto \cite{Tsukamoto:2016qro} identified certain limitations in Bozza's strong deflection limit analysis within ultrastatic spacetimes and subsequently developed a new deflection angle formula for ultrastatic Ellis spacetimes \cite{Tsukamoto:2016qro}. Further refinements to the deflection angle analysis in the Reissner-Nordstr\"{o}m spacetime were made by \cite{Eiroa:2002mk} and Tsukamoto and Gong \cite{Tsukamoto:2016oca}.  
  The work of Virbhadra and Ellis \cite{Virbhadra:1999nm}, Bozza \cite{Bozza:2001xd, Bozza:2002zj, Bozza:2002af, Bozza:2010xqn}, and Tsukamoto \cite{Tsukamoto:2016jzh} has become fundamental in studying strong-field gravitational lensing. These methodologies are extensively used to explore lensing effects in black holes within alternative gravity theories \cite{Ghosh:2020spb, Whisker:2004gq, Eiroa:2005ag, Gyulchev:2006zg, Ghosh:2010uw, Gyulchev:2012ty, Molla:2023hog, Grespan:2023cpa, Kumar:2022fqo, KumarWalia:2022ddq, Kumar:2021cyl, Lu:2021htd, Ali:2021psk, Hsieh:2021scb, Gyulchev:2012ty, Vachher:2024ldc, Molla:2024yde, Molla:2023yxn}, making gravitational lensing a powerful observational tool to distinguish between classical GR and modified theories of gravity.

Recent research has explored quantum-corrected versions of the Kerr black hole within frameworks such as LQG and string theory \cite{Liu:2020ola, Brahma:2020eos, KumarWalia:2022ddq, Islam:2022wck, Afrin:2022ztr, Yang:2022btw, Kumar:2023jgh}. These models introduce parameters that modify the black hole's geometry, particularly near the event horizon. For example, Lewandowski \textit{ et al.} \cite{Lewandowski:2022zce} developed a quantum-corrected model in LQG, altering the Schwarzschild solution to prevent singularities by halting collapse at the Planck scale. This model has been investigated to discuss shadows, photon rings, quasi-normal modes, and gravitational lensing \cite{Yang:2022btw, Gong:2023ghh, Ye:2023qks, Zhao:2024elr}. 

 The Event Horizon Telescope (EHT) released the shadow images of M87* and Sgr A*, which has provided vital insights into black hole physics and offered noteworthy constraints on modified gravity theories. The shadow of the supermassive black holes (SMBHs) M87* \cite{EventHorizonTelescope:2019dse, EventHorizonTelescope:2019pgp, EventHorizonTelescope:2019ggy} and Sgr A* \cite{EventHorizonTelescope:2022exc, EventHorizonTelescope:2022xqj} released by the Event Horizon Telescope (EHT) collaboration have ushered in a new era of black hole physics. These images provide a direct view of supermassive black holes and their surrounding environments.
For M87*, with a distance of 16.8 Mpc and a mass of $(6.5 \pm 0.7) \times 10^9 M_\odot$, the data constrain the size of the compact emission region to an angular diameter of $d_{sh}=42 \pm 3, \mu$as. In 2022, the EHT collaboration also released results for Sgr A* \cite{EventHorizonTelescope:2022exc, EventHorizonTelescope:2022urf, EventHorizonTelescope:2022apq, EventHorizonTelescope:2022wok, EventHorizonTelescope:2022wkp, EventHorizonTelescope:2022xqj}. For Sgr A*, with a mass of $4.0^{+1.1}_{-0.6} \times 10^6 M\odot$ and a distance of 8 kpc, the shadow has an angular diameter of $d_{sh}=48.7 \pm 7,\mu$as. These observations are consistent with the expected appearance of a Kerr black hole \cite{EventHorizonTelescope:2022exc, EventHorizonTelescope:2022xqj}. 
These observations indicated black hole shadows consistent with general relativity's (GR) predictions, but they also offer a stringent test for any deviations introduced, e.g.,  by quantum corrections.   The EHT observations of M87* and Sgr A* provide a unique opportunity to constrain various black hole parameters and evaluate the associated gravitational theories \cite{EventHorizonTelescope:2021dqv,EventHorizonTelescope:2022xqj,Ghosh:2020spb,Afrin:2021imp,KumarWalia:2022aop,Kumar:2022fqo,Islam:2022ybr,Sengo:2022jif}.

These breakthroughs, along with analytical methods for gravitational lensing and shadow analysis \cite{Kumar:2023jgh, Molla:2024yde,Kuang:2022ojj,Molla:2024lpt,Kumar:2023jgh,Kumar:2022fqo,Islam:2022ybr,Kumar:2021cyl,Islam:2021ful,Islam:2021dyk}, offer a unique opportunity to refine our understanding of strong-field gravity and place stringent limits on alternative theories. Integrating observational results from EHT for M87* and Sgr A* with theoretical models deepens our insight into gravity in the Universe's most extreme conditions. Understanding quantum corrections' impact on gravitational lensing is crucial for two reasons. First, strong-gravitational lensing provides direct observational signatures to test deviations from GR, allowing us to probe the photon sphere and relativistic images near black holes. Second, comparing theoretical lensing predictions in quantum-corrected spacetimes with precise measurements from the Event Horizon Telescope (EHT) can constrain the quantum correction parameter $\alpha$. This paper investigates how quantum corrections, specifically those inspired by Loop Quantum Gravity (LQG), affect the lensing properties of rotating black holes. By introducing quantum corrections into the Kerr black hole solution, we obtain the Rotating Quantum-Corrected Black Hole (RQCBH), where $\alpha = 0$ corresponds to the classical Kerr solution. We derive RQCBH using the updated Newman-Janis generating method \cite{Azreg-Ainou:2014pra, Azreg-Ainou:2014aqa}, which is effective for generating rotating metrics from non-rotating seed metrics in various modified gravity theories \cite{Azreg-Ainou:2014pra, Azreg-Ainou:2014aqa, Ghosh:2021clx, Ghosh:2014pba, Mazza:2021rgq,Afrin:2021wlj,Ghosh:2021clx,Walia:2021emv,Kumar:2020yem,Kumar:2020owy,Kumar:2020hgm,Kumar:2019pjp,Kumar:2020ltt,Kumar:2019ohr}, including those inspired by LQG \cite{Brahma:2020eos,Liu:2020ola,Chen:2022nix,Modesto:2008im,Afrin:2022ztr,Islam:2022wck}.

This paper addresses this gap by deriving the strong-gravitational lensing formalism for an RQCBH and comparing the results with those of the Kerr black hole. Our analysis follows the strong deflection limit approach developed by Bozza \cite{Bozza:2002zj}, which has proven effective in describing lensing near compact objects where the gravitational field is strong enough to bend light by angles much more significant than in the weak-field regime. Using this formalism to the RQCBH spacetime, we examine how the quantum parameter $\alpha$ alters key lensing observables, including the deflection angle, photon sphere radius, image magnification, and time delays between relativistic images. By comparing the predicted shadow size and lensing properties of RQCBHs with the EHT results, we can place constraints on the quantum correction parameter $\alpha$ and explore whether quantum gravity effects are detectable with current observational capabilities.

This paper is organized as follows: Section \ref{sec2} presents the rotating quantum-corrected black hole solution and discusses its properties. Section \ref{sec3} presents the formalism for strong gravitational lensing in the RQCBH spacetime and discusses the key observables associated with lensing, including deflection angles, photon sphere radius, and magnification of images. In Section \ref{Sec4}, we numerically analyze the strong lensing observables and time delays for RQCBHs by taking the supermassive black holes Sgr A* and M87* as the lens and compare these results with the Kerr black hole predictions. In Section \ref{Sec5}, we derive constraints on the quantum correction parameter $\alpha$ using the EHT observations. Finally, in Section \ref{Sec6}, we summarize our findings and discuss the implications for quantum gravity theories.

In what follows, we set $G = c = 1$ unless otherwise stated and work with the signature convention $\{-, +, +, +\}$. 

\section{Rotating Quantum-Corrected Black Holes }\label{sec2}
 {We are interested in the quantum-corrected black hole metric achieved through a pseudo-static setup, where the Killing vector $\partial_t$ is space-like inside the black hole. Notably, this metric is obtained by applying junction conditions across the collapsing boundary without explicitly using the equations of motion \cite{Lewandowski:2022zce}, with the black hole metric as follows}
\begin{equation}
ds^2=-f\left(r\right)dt^2+\frac{1}{f\left(r\right)}dr^2+r^2\left(d\theta^2+{sin}^2\theta d\phi^2\right),
\label{1}
\end{equation}
where
\begin{equation}
f\left(r\right)=1-\frac{2M}{r}+\frac{\alpha{M}^2}{r^4}.
\label{2}
\end{equation}
Here, $\alpha=16\sqrt3\gamma^3$ is the quantum correction parameter \cite{Lewandowski:2022zce}. The parameter $M$ coincides with the ADM mass of the metric \ref{1}. There is a lower bound for mass $M$ given by
\begin{equation}
M_{min}=\frac{4\sqrt{\alpha}}{3\sqrt{3}G}=\frac{16\gamma\sqrt{\gamma\pi}}{3\sqrt[4]{3}}\approx0.8314.
\label{3}
\end{equation} such that for $M<M_{min}$, metric (\ref{1}) does not admit any horizon, or for $M>M_{min}$ there are two horizon viz Cauchy horizon($r_-$) and Event horizon ($r_+$). We shall focus on the later case in this paper. 
The minimal mass is of the order of Planck mass whose value depends on the value of Barbero-Immirzi parameter $\gamma\approx0.2375$ \cite{Lewandowski:2022zce, Ye:2023qks}
When $M>M_{min}$ or $1/2<\beta<1$, the metric admits two horizons viz \cite{Lewandowski:2022zce}
\begin{equation}
    r_\pm=\frac{\beta(1\pm\sqrt{2\beta-1}}{\sqrt{(1+\beta)(1-\beta)^3}}\sqrt{\alpha}
\end{equation}where
\begin{equation}
   \mathbf{ M^2=\frac{4\beta^4}{(1-\beta^2)^3}\alpha}
\end{equation}
By examining the metric (\ref{2}), we observe that  $ \lim\limits_{r \to \infty} f\left(r\right)=1$, indicating the  spacetime to be asymptotically flat. However, non-rotating black holes, like those described by the Schwarzschild metric, are difficult to observe directly, as spin plays a fundamental role in astrophysical processes. The Kerr metric, which describes a rotating black hole, is one of the most significant solutions of GR resulting from gravitational collapse. This completion motivates the search for an axisymmetric generalization of the quantum-corrected black hole metric or a Kerr-like rotating metric. The goal is to find a rotating quantum-corrected black hole metric (RQCBH), which could then be tested against EHT observations via gravitational lensing to understand better the role of quantum effects in rotating black holes.

We derive RQCBH using the updated Newman-Janis generating method \cite{Azreg-Ainou:2014pra, Azreg-Ainou:2014aqa}, which has been successful in generating imperfect fluid rotating solutions in Boyer-Lindquist coordinates from spherically symmetric static solutions, and it can also produce generic rotating regular black hole solution.
 In the Boyer-Lindquist coordinates, the RQCBH metric can be written as \cite{Kumar:2022vfg, Islam:2022wck}.
\begin{eqnarray}\label{metric3}
ds^2 &=& -\left[1-\frac{1}{\rho^2}\left(2Mr- \frac{\alpha M^2}{r^2}\right)\right] dt^2+ \frac{\rho^2}{\Delta} dr^2 +\rho^2 d\theta^2  \nonumber\\ && - \frac{2a}{\rho^2}\left(2Mr-\frac{\alpha M^2}{r^2}\right) \sin^2\theta dtd\phi+ \frac{\Sigma\sin^2\theta~}{\rho^2} d\phi^2\nonumber\\
\end{eqnarray}
where
\begin{eqnarray}
\rho^2 = r^2 +a^2\cos^2\theta,\nonumber \\
\mathbf{\Delta = r^2 +a^2 -2 Mr+ \frac{\alpha M^2}{r^2}},\nonumber \\
\Sigma = (r^2 +a^2)^2-a^2 \Delta \sin^2\theta.
\end{eqnarray}
The rotating metric provides important insights into the behavior of rotating black holes in the presence of quantum effects. It is important to note that in the limit $a \to 0$, the RQCBH reduces to a spherical quantum-corrected black hole (QCB) \cite{Lewandowski:2022zce} and for $\alpha \to 0$ (Eq.~(\ref{metric3})) encompasses the Kerr spacetime \cite{Kerr:1963ud}. Moreover, when $a=M= \alpha= 0$, then Eq.~(\ref{metric3}) describes flat spacetime.

\begin{figure}[!tbh] 
	\begin{centering}
		    \includegraphics[width=\columnwidth]{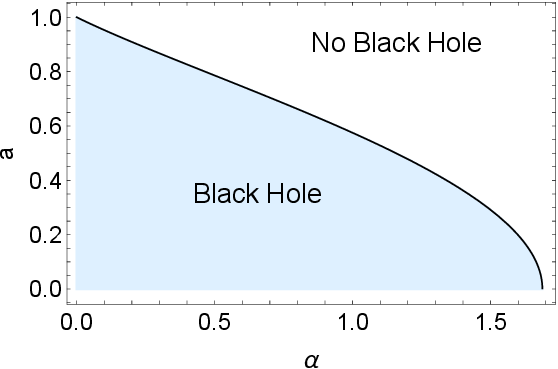}
	\end{centering}
	\caption{Parameter space ($a$, $\alpha$) for RQCBH spacetime. The black solid line corresponds to the extremal black holes with degenerate horizons}\label{plot1}
\end{figure}

We solve the Equation $\Delta(r)=0$, and we can find the radii of the RQCBH horizons, which is a coordinate singularity of metric (\ref{metric3}). Depending on the values of $a$ and $\alpha$, there may be up to two positive roots. The quantum correction parameter, $\alpha$, must be within the range $27/16 \ge \alpha \ge 0$ as shown in the parameter space plot for RQCBH in Fig.~\ref{plot1}. In the blue region of the parameter space, the black hole admits two positive roots, corresponding to the RQCBHs with Cauchy and event horizons. At $\alpha=\alpha_c$, denoted by the black solid line, the Equation $\Delta(r) = 0$ has a double root for a given $a$, corresponding to an extremal RQCBH with degenerate horizons. For $\alpha>\alpha_c$, the Equation $\Delta(r) = 0$ has no positive roots corresponding to no-horizon (NH) spacetimes. 

\section{Strong Gravitational Lensing by RQCBH}\label{sec3}
\begin{figure*}[t!] 
	\begin{centering}
		\begin{tabular}{c c}
		    \includegraphics[width=\columnwidth]{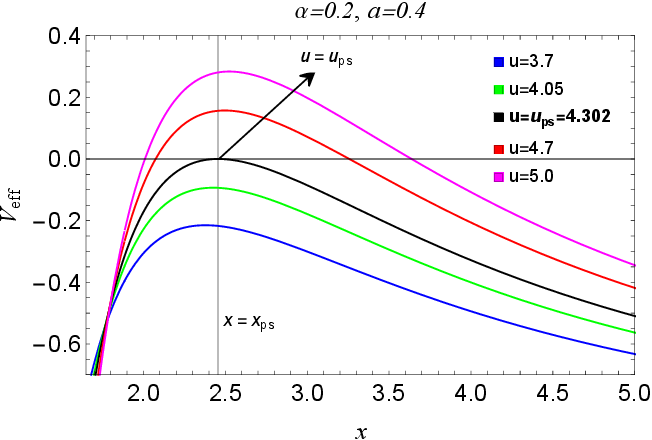}\hspace{0.5cm}
		    \includegraphics[width=\columnwidth]{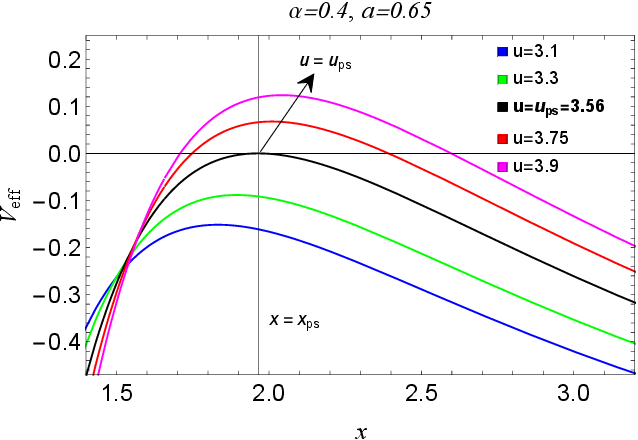}
			\end{tabular}
	\end{centering}
	\caption{The radial effective potential for RQCBHs as a function of radial distance $x$ for different values of impact parameters $u$. Photons with $u=u_{ps}$ represented by the solid black curves, orbit around the black hole in unstable circular paths with radius $x=x_{ps}$. Photons with impact parameters greater than $u_{ps}$ ( $u > u_{ps}$) make multiple loops around the black hole before scattering to infinity. These photons form the strong gravitational lensing image of the source. On the other hand, photons with impact parameters less than $u_{ps}$($u < u_{ps}$) are captured by the black hole. }\label{plot2}
\end{figure*}

By considering the light rays in the equatorial plane ($\theta= \pi/2$) and re-scaling the quantities $r, a, \alpha$,  and $t$ in units of $M$ \cite{Bozza:2002zj} such that
\[
r/M \to x, ~~~~a/M \to a,~~~~\alpha/M=\alpha  ~~~~\text{and} ~~~t/M \to t, 
\]
We rewrite the RQCBH metric (\ref{metric3}) on the equatorial plane as 
\begin{eqnarray}\label{NSR}
\mathrm{ds^2}=-A(x)dt^2+B(x) dx^2 +C(x)d\phi^2-D(x)dt\,d\phi,\nonumber\\
\end{eqnarray}
where
\begin{eqnarray}\label{compo}
A(x) &=& 1-\frac{1}{\rho^2}\left(2x-\frac{\alpha}{x^2}\right),~~~~
B(x)=\frac{{\rho}^2}{\Delta}, \nonumber\\
C(x) &=& \frac{\Sigma}{{\rho}^2},~~~~
D(x) = \frac{2a}{{\rho}^2}\left(2Mx-\frac{\alpha}{x^2}\right) ,    
\end{eqnarray}
with ${\rho}^2 = x^2 $ and $\mathcal{A} =(x^2 +a^2)^2-a^2\Delta $. 
We use the Hamilton-Jacobi method to examine the motion of the photon trajectory around the black hole. The null geodesics that describe photon orbits around black holes can be described by the Lagrangian $\mathcal{L} =\frac{1}{2}g_{\mu\nu}\dot x^{\mu}\dot x^{\nu} = 0$, where the dot means the derivative with respect to the affine parameter $\lambda$. Now, we introduce the Hamilton-Jacobi equation
\begin{equation}
    \mathcal{H}=-\frac{\partial S}{\partial\lambda}=\frac{1}{2}g_{\mu\nu}\frac{\partial S}{\partial x^{\mu}}\frac{\partial S}{\partial x^{\nu}}\label{ham}
\end{equation}
where $\mathcal{H}$ and $S$ are the canonical Hamiltonian and the Jacobi action. Two linearly independent killing vectors,  associated with the time translation and rotational invariance,  are admitted by the black hole metric- The total energy $\mathcal{E}$ and the angular momentum $\mathcal{L}$ (\ref{metric3})
\begin{equation}
    \mathcal{E}=-p_t=-(g_{tt}\dot t+g_{t\phi}\dot \phi),\text{and}~~\mathcal{L}=p_\phi=g_{\phi t}\dot t-g_{\phi\phi}\dot \phi.
    \label{conserve}
\end{equation}
Using the equation of motion for the photons and the two conserved quantities, we establish a relationship between the effective potential $V_{\text{eff}}$ and impact parameter $u$. The relationship between $u(=\frac{\mathcal{L}}{\mathcal{E}})$. The effective potential for the radial motion is given as:
\begin{equation}
    V_{eff}=-\frac{4(Au^2+Du-C)}{B(4AC+D^2)}\label{veff}
\end{equation}
which describes different photon orbits around the black hole. A photon emitted from a source travels toward the observer and gets deflected by a black hole when it reaches distance $x_0$ due to the gravitational field of the black hole. The impact parameter $u$ is defined as the perpendicular distance from the center of mass of the black hole to the photon's initial direction, influencing the radial motion and determining the closest approach distance $x_0$. It turns out that photon rays exist in the region where $V_{eff}\le \mathcal{E}^2$. We can define the unstable photon orbit $x_{ps}$ satisfying $V_{eff}(x)=V_{eff}'(x)=0$ and $V_{eff}''<0$. Photons move along these unstable circular photon orbits $x_{ps}$ until small radial perturbations drive these photons either into the black hole or toward spatial infinity. The gravitationally-lensed projection of this photon sphere on the observer's image plane creates the black hole shadow. The shape and size of the black hole shadow can provide critical insights into the characteristics of the gravitational field near the horizon, making it a useful tool for testing the signatures of strong gravitational lensing. As shown in Fig.~\ref{plot2}, the photon propagates only in the region where $V_{eff}\leq\mathcal{E}^2$. At the closest approach distance $x_0$ or $V_{eff}=0$, the impact parameter $u$ reads
\begin{equation}
   u(x_0)=\frac{-D_0+\sqrt{4A_0C_0+D_0^2}}{2A_0}, \label{impact}
\end{equation}
 The condition for the solution of an unstable circular photon orbit $x_{ps}$ in terms of metric components is given as,

\begin{eqnarray}
\label{ps}
A(x)C'(x)-A'(x)C(x) + u(A'(x)D(x) - A(x)D'(x)) &=& 0,\nonumber\\   
\end{eqnarray}

which implies 
\begin{eqnarray}\label{ps1}
\frac{1}{(x - 2)x^3 + x\alpha} \Bigg[2 ax\left (x^3 -2\alpha\right)\sqrt {a^2 + (x - 2)x+\frac{\alpha}{x^2}} \nonumber\\ 
- 2 a^2 (x^5 - 2 x^2 \alpha)+ \left(( x-2) x^3 + \alpha) ((x-3) x^3 + 3 \alpha\right)\Bigg] &=& 0\nonumber\\
\end{eqnarray}

\begin{figure*} 
	\begin{centering}
		\begin{tabular}{p{9cm} p{9cm}}
		    \includegraphics[width=\columnwidth]{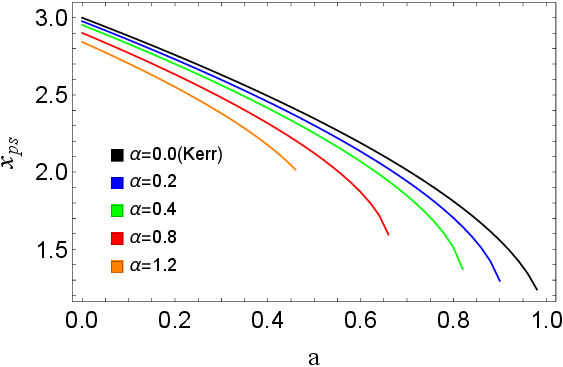}&
		    \includegraphics[width=\columnwidth]{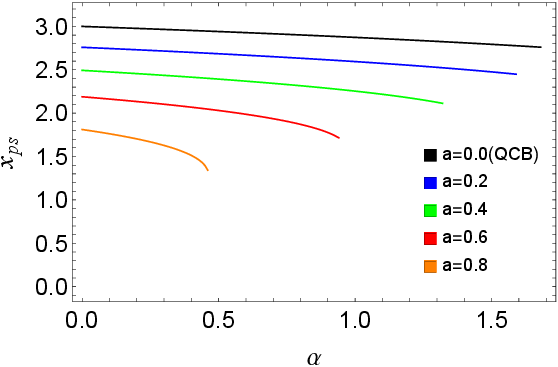}
			\end{tabular}
	\end{centering}
	\caption{Behavior of the photon sphere radius $x_{\text{ps}}$ with respect to the parameter $a$ (left) for different $\alpha$ and with respect to the parameter $\alpha$ (right) for different $a$ for RQCBH spacetime.}\label{plot4}
\end{figure*}

\begin{figure*}
	\begin{centering}
		\begin{tabular}{p{9cm} p{9cm}}
            \includegraphics[width=\columnwidth]{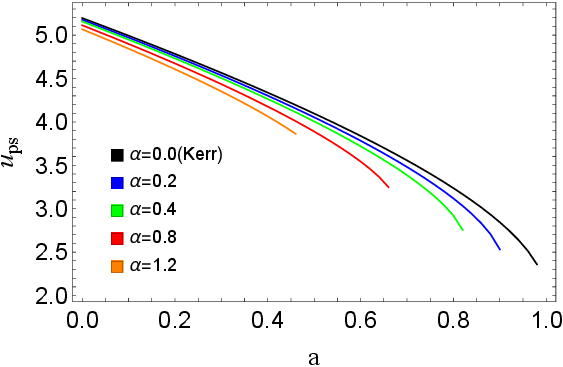}&
		    \includegraphics[width=\columnwidth]{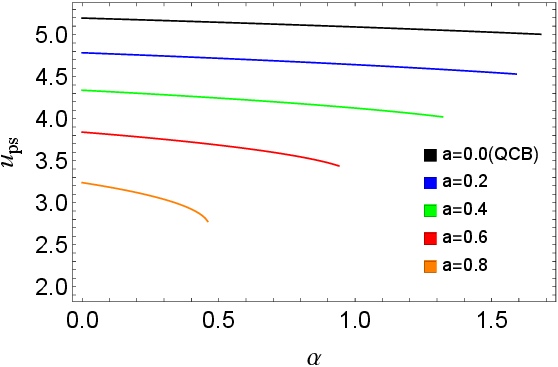} 
			\end{tabular}
	\end{centering}
	\caption{Behavior of the critical impact parameter $u_{\text{ps}}$ with respect to the parameter $a$ (Left) for different $\alpha$ and with respect to the parameter $\alpha$ (Right) for different $a$ for RQCBH spacetime.}\label{plot5}
\end{figure*}

The unstable photon orbit radius is the largest root of Eq.~\ref{ps1} \cite{Claudel:2000yi,Virbhadra:2002ju}. 
 Figure~\ref{plot4} shows a decrease in the critical photon radius $x_{ps}$ with both rotation parameter $a$ and quantum parameter $\alpha$ which shows that the $x_{ps}$ of Kerr black holes is greater than that of RQCBHs. Figure~\ref{plot4} also suggests that the photons forming prograde orbits can get closer to the black hole than the photons forming retrograde orbits. When the closest approach distance $x_0\to x_{ps}$, the impact parameter $u$ is called the critical impact parameter $u_{ps}$ given as,
 \begin{equation}
   u_{ps}=  \frac{a ( \alpha-2 x_{ps}^3) + 
 x^4 \sqrt{a^2 - 2  x_{ps} +  x_{ps}^2 +\frac{\alpha}{ x_{ps}^2}}}  { x_{ps}^4-2  x_{ps}^3 + \alpha}
 \end{equation}
 Figure~\ref{plot5} shows the variation of the critical impact parameter with both $a$ and $\alpha$ and found that it varies similarly to unstable photon orbit radius.


\subsection{Deflection angle of strong-gravitational lensing}
\begin{figure*}[t] 
	\begin{centering}
		\begin{tabular}{cc}
		    \includegraphics[width=\columnwidth]{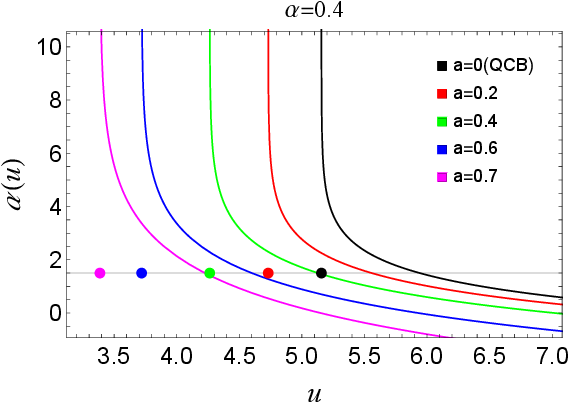}\hspace{0.5cm}
		    \includegraphics[width=\columnwidth]{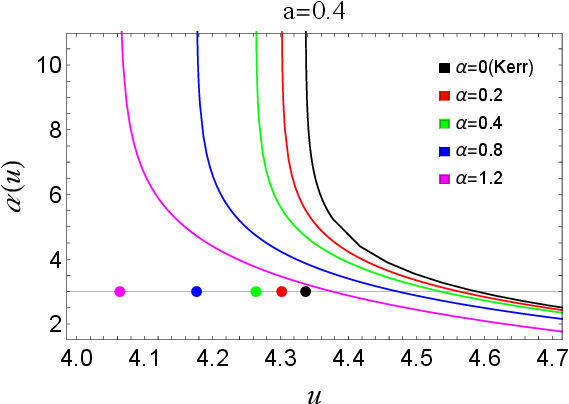}
			\end{tabular}
	\end{centering}
	\caption{Variation of deflection angle, in strong field limit, for RQCBH spacetime as a function of the impact parameter $u$ for different values of the parameters $a$ (Left) and $\alpha$ (Right) taking one of the values of the parameter as constant. The dots on the horizontal axis represent the values of the critical impact parameter $u_{ps}$ at which the deflection angle diverges. }\label{plot6}
\end{figure*}

In this section, we shall study the strong-gravitational lensing by RQCBH and explore the effect of its parameters on the lensing observables under the strong-field limit.
In SDL, when the closest distance approaches $x_0\to x_{ps}$, the deflection angle increases and eventually exceeds 2$\pi$ radians and becomes unboundedly large at $x_0 = x_{ps}$. 
The deflection angle by the rotating black hole for the photon moving in the equatorial plane as a function of the closest distance $x_0$ can be obtained as: \cite{Islam:2021ful, Islam:2021dyk, Bozza:2002zj, Virbhadra:1998dy}
\begin{eqnarray}\label{bending2}
\alpha_{D}(x_0) &=&  I(x_0) -\pi
\end{eqnarray}
where
\begin{eqnarray}
    &&I(x_0)=2\int_{x_0}^{\infty}\frac{d\phi}{dx}dx
   \nonumber\\&=&2\int_{x_0}^{\infty}\frac{\sqrt{A_0 B }\left(2Au+ D\right)}{\sqrt{4AC+D^2}\sqrt{A_0 C-A C_0+u\left(AD_0-A_0D\right)}}dx,
\nonumber\\
\end{eqnarray}
Since the integral cannot be solved explicitly, therefore the integral expanded near the unstable photon sphere radius \cite{Virbhadra:1999nm,Claudel:2000yi,Bozza:2002zj} by defining a new variable $z=1-x_0/x$ in SDL \cite{Islam:2022ybr,Zhang:2017vap}. The analytic expression of the deflection angle for the spacetime (\ref{ps1}) as a function of impact parameter ($u\approx\theta D_{OL}$) is given by \cite{Bozza:2002zj,Kumar:2020sag,Islam:2020xmy}
\begin{eqnarray}\label{def4}
\alpha_{D}(u) &=& \bar{a} \log\left(\frac{u}{u_{ps}} -1\right) + \bar{b} + \mathcal{O}(u-u_{ps}),  
\end{eqnarray}  

where $\bar{a}$, $\bar{b}$ are the strong lensing coefficients. Detailed calculations can be found in \cite{Bozza:2002zj,Kumar:2020sag,Islam:2020xmy}. 
We compare and analyze the results of the deflection angle of strong gravitational lensing by RQCBH with the analogous results of the Kerr black hole. The effect of the RQCBH parameter $\alpha$ and the spin $a$ can be seen in Fig.~\ref{plot6}. The deflection angle diverges at larger $u_{ps}$ for smaller values of the parameters $\alpha$ and $a$. In comparison, we see that the deflection angle for Kerr black holes is larger than RQCBHs for a fixed value of impact parameter $u$ (c.f. Fig.~\ref{plot6}).

\subsection{Lensing Observables}

 Suppose the source and observer are located far from the lens, nearly aligned, and located in a flat spacetime \cite{Bozza:2002zj, Bozza:2008ev}. In that case, the lens equation provides a geometric framework to describe the relationships between the observer, the lens, and the light source.
\begin{eqnarray}\label{lenseq}
\beta &=& \theta -\frac{D_{LS}}{D_{OL}+D_{LS}} \Delta\alpha _n,
\end{eqnarray}
where $\Delta \alpha_{n}=\alpha(\theta)-2n\pi$ is the extra deflection angle of a photon looping over $2 n \pi$ and $n$ being an integer $n \in N$, and $0 < \Delta \alpha_n \ll 1$. In addition, $\beta$ and $\theta$ are the angular positions of the source and the image from the optical axis; the projected distances of the lens(L) from the source(S) and the observer(O) to the lens(L) are $D_{\mathrm{LS}}$ and $D_{\mathrm{OS}}\approx D_{OL}+D_{LS}$. Using this formula, we can estimate the distance of the black hole.

Now we estimate the observables for the strong gravitational lensing by the RQCBH spacetime as in \cite{Islam:2021dyk,Ghosh:2020spb,Bozza:2002af,Bozza:2002zj}. Using the lens Eq.~(\ref{lenseq}) and Eq.~(\ref{def4}) and the relation $u\approx\theta D_{OL}$, the angular separation between the lens and the $n^{th}$ relativistic image in terms of strong lensing coefficients is given by~\cite{Bozza:2002af}  
\begin{eqnarray}\label{angpos}
\theta_n &=& \theta_n{^0} +\frac{D_{OL}+D_{LS}}{D_{LS}}\frac{u_{ps}e_n}{\bar{a} D_{OL}}(\beta-\theta_n{^0}),
\end{eqnarray} 
where 
\begin{eqnarray}
\theta_n{^0} &=& \frac{u_{ps}}{D_{OL}}(1+e_n), \label{postion}\\
e_n &=& \text{exp}\left({\frac{\bar{b}}{\bar{a}}-\frac{2n\pi}{\bar{a}}}\right).
\end{eqnarray}
Here $\theta_n{^0}$ is the angular position of the image when a photon encircles complete $2n\pi$ and the second term in Eq.~(\ref{angpos}) is the correction term for $\theta_n{^0} \gg \Delta\theta_n $ \cite{Bozza:2002zj}. When the particular orientation of source, lens and observer is perfectly aligned ($\beta = 0$), then Eq.~(\ref{angpos}) gives the angular radius of the Einstein rings for spherically symmetric cases \cite{Einstein:1936llh,Liebes:1964zz,Mellier:1998pk,Bartelmann:1999yn,Schmidt:2008hc}. Note that Einstein rings are called relativistic Einstein rings when the deflection angle is greater than $2\pi$. Taking $D_{OS}=2D_{OL}$ and $D_{OL}\gg u_{ps}$, then Eq.~(\ref{angpos}) reduces to
\begin{equation}
  \theta_n^E = \frac{u_{ps}}{D_{OL}}(1+e_n)  
\end{equation}
Einstein rings for supermassive black holes depend on the parameters of RQCBHs, with the outermost ring being the Kerr with $\alpha=0$. The Einstein ring's size decreases as the quantum-corrected parameter $\alpha$ increases. 

\begin{table*}[htb!]
\resizebox{18cm}{!}{
 \begin{centering}	
	\begin{tabular}{p{2cm} p{2cm} p{2cm} p{2cm} p{2cm} p{2cm} p{2cm}}
\hline\hline
\multicolumn{2}{c}{}&
\multicolumn{2}{c}{Sgr A*}&
\multicolumn{2}{c}{M87*}\\
{$a$ } & {$\alpha$}& {$\theta_{\infty}$($\mu$as)} & {$s$ ($\mu$as)} & {$\theta_{\infty}$($\mu$as)}  & {$s$ ($\mu$as) } & {$r_{mag}$} \\ \hline
\hline
\multirow{7}{*}{0.0}&0.0& 26.330 & 0.033 & 19.782 &0.025& 6.822\\
& 0.2 & 26.23 & 0.035 & 19.707 & 0.026 & 6.743\\
& 0.4 & 26.13& 0.0373 & 19.63 & 0.028 & 6.658\\
& 0.6 & 26.02 & 0.040 & 19.55 & 0.030 & 6.566\\
& 0.8 & 25.91 & 0.0431 & 19.46 & 0.032 &6.466\\
& 0.9 & 25.85 & 0.0449 &  19.42 & 0.033 & 6.412\\
\hline
\multirow{7}{*}{0.4}& 0.0 & 21.977 & 0.087 & 16.511& 0.065 & 5.587\\
& 0.2 &  21.798& 0.0968 & 16.37 & 0.072 & 5.435 \\
& 0.4 & 21.606& 0.109 & 16.23 &0.082 & 5.26\\ 
& 0.6 & 21.39 &0.126 & 16.07 & 0.095 & 5.056 \\ 
& 0.8 & 21.163& 0.148 & 15.90 & 0.11 & 4.806 \\   
\hline
\multirow{3}{*}{0.8}& 0.0 & 16.404 & 0.376 & 12.325 & 0.282 & 3.561 \\
& 0.2 & 15.803& 0.522 & 11.87 &  0.392 & 2.988\\ 
& 0.4 & 14.803& 0.792 & 11.12 & 0.595 & 1.817 \\

		\hline\hline
	\end{tabular}
\end{centering}
}	
	\caption{Estimates for the lensing observables by considering supermassive black holes at the center of nearby galaxies as RQCBHs. We measure the quantities $a$ and $\alpha$ in units of the mass of the black hole $M$.  }\label{table1}
\end{table*}

\begin{table*}[htb!]
\resizebox{18cm}{!}{
 \begin{centering}	
	\begin{tabular}{p{2cm} p{2cm} p{2cm} p{2cm} p{2cm} p{2cm} p{2cm}}
\hline\hline
\multicolumn{2}{c}{}&
\multicolumn{2}{c}{Sgr A*}&
\multicolumn{2}{c}{M87*}\\
{$a$ } & {$\alpha$}& {$\Delta\theta_{\infty}$($\mu$as)} & {$\Delta s$ ($\mu$as)} & {$\Delta\theta_{\infty}$($\mu$as)}  & {$\Delta s$ ($\mu$as) } & {$\Delta r_{mag}$} \\ \hline
\hline
\multirow{7}{*}{0.4}& 0.2 &  -0.179& 0.0098 & -0.141 & 0.007& -0.152 \\
& 0.4 & -0.371& 0.022 & -0.281 &0.017 & -0.327\\ 
& 0.6 & -0.587 &0.039 & -0.441 & 0.03 & -0.531 \\ 
& 0.8 & -0.814& 0.061 & -0.611 & 0.045 & -0.781 \\   
\hline
\multirow{2}{*}{0.8}& 0.2 & -0.601& 0.146 & -0.455 &  0.11 & -0.573\\ 
& 0.4 & -1.601& 0.416 & -1.205 & 0.313 & -1.714 \\

		\hline\hline
	\end{tabular}
\end{centering}
}	
	\caption{ Deviation of the lensing observables of RQCBHs from Kerr black hole for supermassive black holes Sgr A* and M87* for $a = 0.40$ and $\alpha = 0.80$. Note that $\Delta O=O(RQCBH)-O(Kerr) $ }\label{tab2}
\end{table*}

The magnification, for the $n-$loop images, is evaluated as the quotient of the solid angles subtended by the $n^{th}$ image and the source $\mu=\sin\theta d\theta/\sin\beta d\beta$, which can be deduced as \cite{Bozza:2002af,Bozza:2002zj}
\begin{eqnarray}\label{mag}
\mu_n &=&  \frac{1}{\beta} \Bigg[\frac{u_{ps}}{D_{OL}}(1+e_n) \Bigg(\frac{D_{OL}+D_{LS}}{D_{LS}}\frac{u_{ps}e_n}{D_{OL} \bar{a}}  \Bigg)\Bigg].
\end{eqnarray}
\begin{figure*}
	\begin{centering}
		\begin{tabular}{c c}
		    \includegraphics[width=\columnwidth]{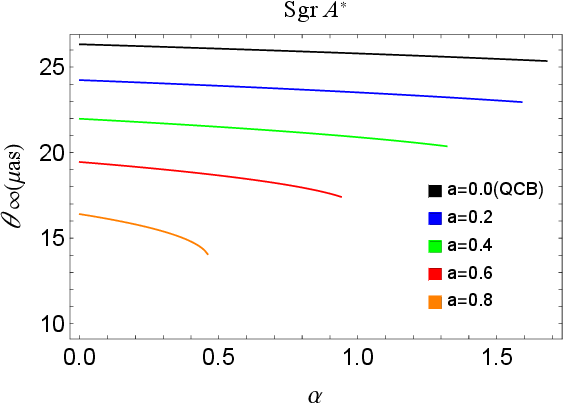}&
            \includegraphics[width=\columnwidth]{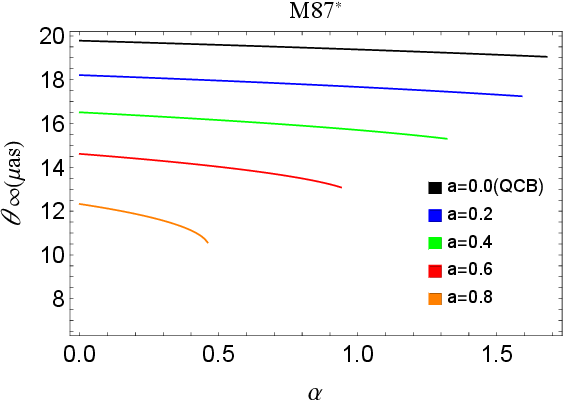}\\
			\includegraphics[width=\columnwidth]{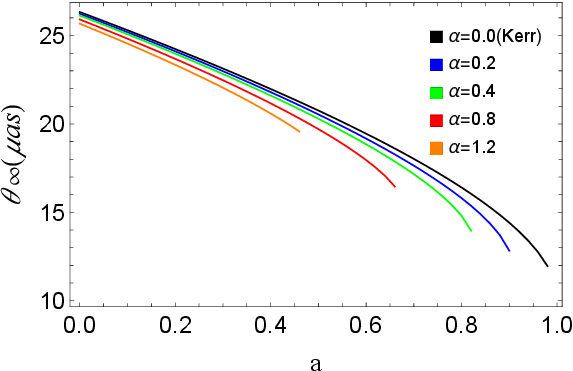}&
			\includegraphics[width=\columnwidth]{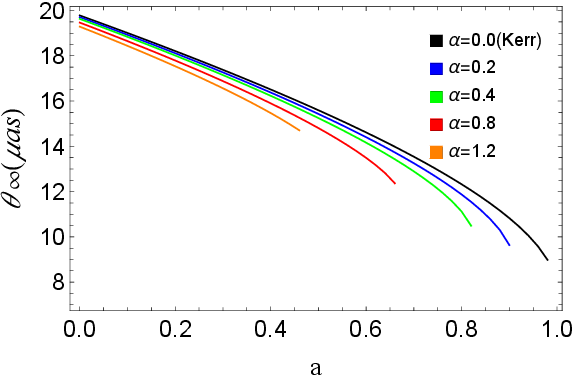}\\
			\end{tabular}
	\end{centering}
	\caption{Behavior of strong lensing observables $\theta_\infty$ in strong field limit, as a function of the parameters $a$ and $\alpha$ by considering that the spacetime around the compact objects at the centers of Sgr A*(Left panel) and M87*(Right panel) is RQCBH spacetime.} \label{plottheta}		
\end{figure*}
\begin{figure*}
	\begin{centering}
		\begin{tabular}{c c}
		    \includegraphics[width=\columnwidth]{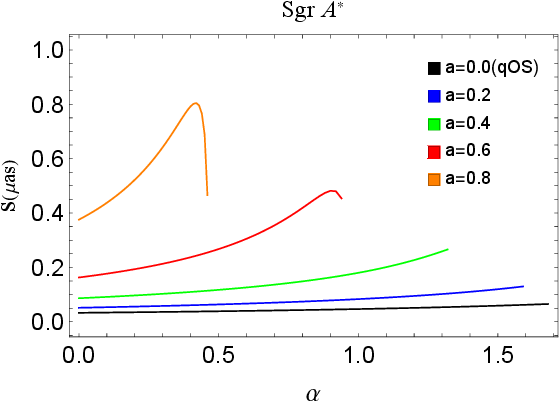}&
            \includegraphics[width=\columnwidth]{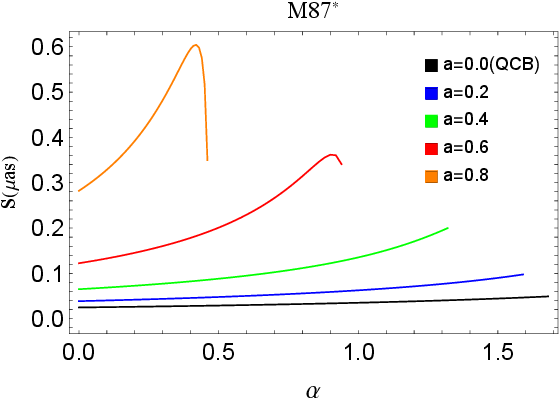}\\
			\includegraphics[width=\columnwidth]{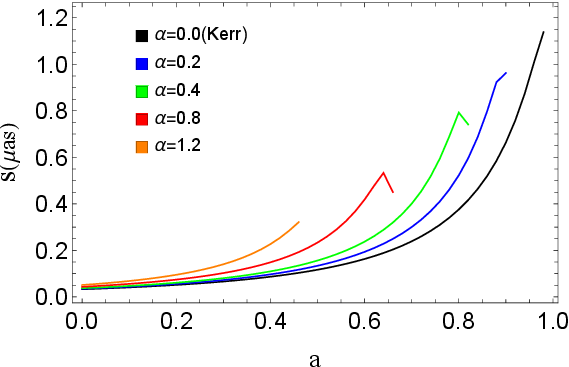}&
			\includegraphics[width=\columnwidth]{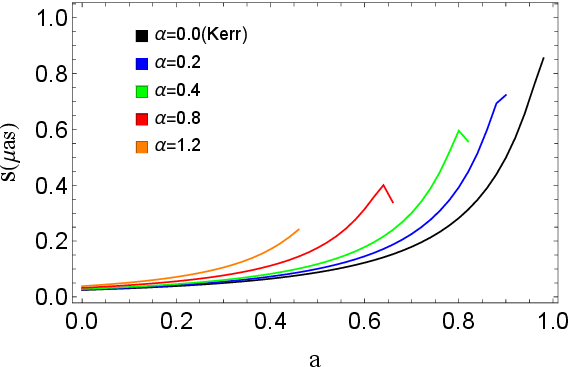}\\
			\end{tabular}
	\end{centering}
	\caption{Behavior of strong lensing observables $s$ in strong field limit, as a function of the parameters $a$ and $\alpha$ by considering that the spacetime around the compact objects at the centers of Sgr A*(Left panel) and M87*(Right panel) is RQCBH spacetime.} \label{plotseperation}		
\end{figure*}

\begin{figure*}
	\begin{centering}
		\begin{tabular}{cc}
		    \includegraphics[width=\columnwidth]{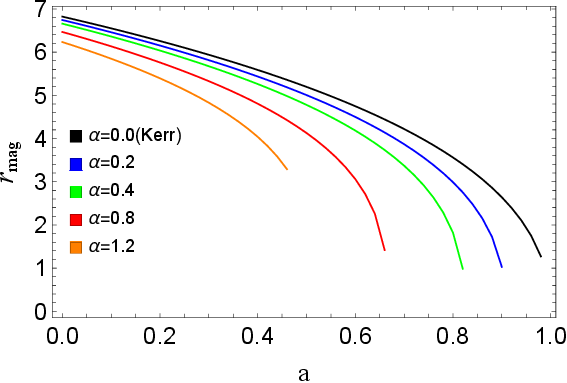}&
		    \includegraphics[width=\columnwidth]{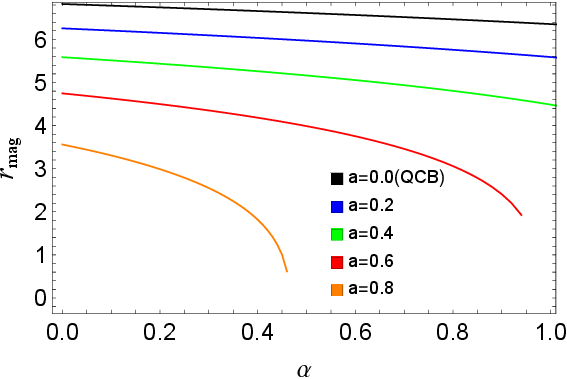}
			\end{tabular}
	\end{centering}
	\caption{Behavior of strong lensing observable $r_{\text{mag}}$, for RQCBH spacetime as a function of the parameters $a$ (Left) and $\alpha$ (Right). Note that $r_{mag}$ is independent of the black hole's mass or distance from the observer.}\label{plotrmag}
\end{figure*}
\begin{figure*}
	\begin{centering}
		\begin{tabular}{c c}
		    \includegraphics[width=\columnwidth]{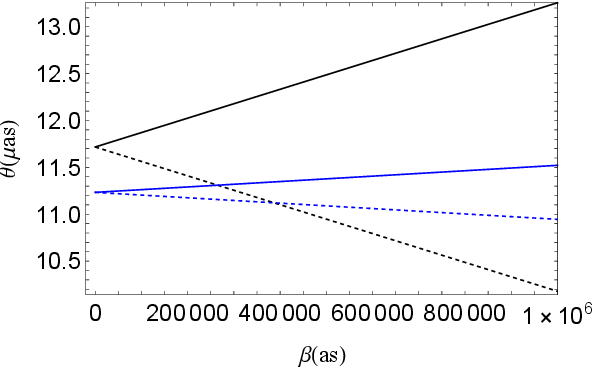}&
			\includegraphics[width=\columnwidth]{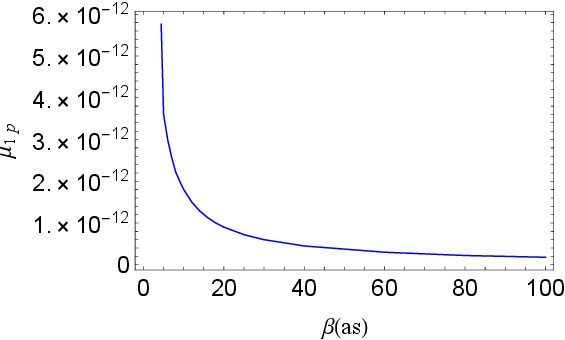}\\
  
			\end{tabular}
	\end{centering}
	\caption{Comparion between the angular position of first primary $\theta_{p}$ (solid line) and secondary image $\theta_{s}$ (dashed line) of first-order(black line) and second-order(blue line) image with position of the source ($\beta$) for M87* (\textit{Left}). The absolute magnification of the first primary image is plotted against the angular source position ($\beta$) for $D_{OS}=2D_{OL}$ (\textit{Right}). } \label{comparion}		
\end{figure*}
 From Eq.~\ref{mag} we see that in the limit $\beta \to 0$, $\mu_n\to\infty$ implying that in case of perfect alignment, the magnification of the images is maximum.
The magnification $\mu_n$ is also inversely proportional to $D_{OL}^2$, so the images are faint and further decrease
with $n$, resulting in images of higher order becoming less visible. So we can see that the brightness of the first image is dominant over all the other relativistic images. As a result, we consider the case where $\theta_{1}$ can be separated as an outermost single image and the remaining images are packed together at $\theta_{\infty}$. Hence we define three characteristic observables~ \cite{Bozza:2002zj} as 
\begin{eqnarray}
\theta_\infty &=& \frac{u_{ps}}{D_{OL}},\label{theta}\\
s &=& \theta_1-\theta_\infty \approx \theta_\infty ~\text{exp}\left({\frac{\bar{b}}{\bar{a}}-\frac{2\pi}{\bar{a}}}\right),\label{sep}\\
r_{\text{mag}} &=& \frac{\mu_1}{\sum{_{n=2}^\infty}\mu_n } \approx 
\frac{5 \pi}{\bar{a}~\text{log}(10)}\label{mag1}.
\end{eqnarray} 
In the above expression, $s$ is the angular separation between $\theta_{1}$ and $\theta_{\infty}$, $r_{\text{mag}}$ is the ratio of the flux of the first image and all the other images.
Note that the observable $r_{\text{mag}}$ does not depend on the distance between the observer and the black hole(lens) $D_{OL}$.

In SDL, multiple relativistic images are created when the deflection angle exceeds 2$\pi$. Consequently, the travel time of light along divergent paths, corresponding to distinct images, is also inherently varied, resulting in a time delay. The time delay between the $k-th$ and $l-th$ images reads as \cite{Bozza:2003cp}
\begin{eqnarray}
\Delta T_{k,l} \approx 2\pi(k-l) u_{ps},
\end{eqnarray}
the time delay between the first and second relativistic image $\Delta T_{2,1}$ when the images are on the same side of the lens can be given as
\begin{eqnarray}
\Delta T_{2,1} = 2\pi u_{ps}=2\pi D_{OL}\theta_\infty,\label{timed}
\end{eqnarray}
Using Eq.~\ref{timed} we compared the time delay between the supermassive black holes at the centre of several galaxies as Kerr and RQCBH spacetime (see Table~\ref{table3} )
\begin{table*}[tbh!]
	\begin{ruledtabular}
		\begin{tabular}{c c c c c c}  
				Galaxy   &           $M( M_{\odot})$      &          $D_{OL}$ (Mpc)   &     $M/D_{OL}$ & $\Delta T^s_{2,1}(\text{Kerr})$&$\Delta T^s_{2,1}(\text{RQCBH})$           \\
			\hline
			Milky Way& $  \mathbf{4\times 10^6}	 $ & $0.0083 $ &       $2.471\times 10^{-11}$ &$\mathbf{10.6947}$  & $\mathbf{6.0127}$     \\
             M87&$ 6.5\times 10^{9} $&$ 16.68 $
&$1.758\times 10^{-11}$& $20488.6 $ &  $18489.$\\			
		
			 NGC 4472 &$ 2.54\times 10^{9} $&$ 16.72 $
&$7.246\times 10^{-12}$& $8461.98$ &  $7636.09$\\
			
			 NGC 1332 &$ 1.47\times 10^{9} $&$22.66  $
&$3.094\times 10^{-12}$& $4897.29$ & $4419.312$ \\
		
			 NGC 4374 &$ 9.25\times 10^{8} $&$ 18.51 $
&$2.383\times 10^{-12}$& $3081.63$ &  $2780.86$\\
			
			NGC 1399&$ 8.81\times 10^{8} $&$ 20.85 $
&$2.015\times 10^{-12}$& $2935.04$ &  $2648.58$\\
			 
			  NGC 3379 &$ 4.16\times 10^{8} $&$10.70$
&$1.854\times 10^{-12}$& $1385.9$ &  $1250.635$\\
			
			 NGC 4486B &$ 6\times 10^{8} $&$ 16.26 $
&$1.760\times 10^{-12}$ & $1998.89$ &  $1803.8$\\
		
			 NGC 1374 &$ 5.90\times 10^{8} $&$ 19.57 $ &$1.438\times 10^{-12}$& $1965.58$ &  $1773.736$\\
			    
			NGC 4649&$ 4.72\times 10^{9} $&$ 16.46 $
&$1.367\times 10^{-12}$& $15724.6$ &  $ 14189.9 $\\
		
			NGC 3608 &$  4.65\times 10^{8}  $&$ 22.75  $ &$9.750\times 10^{-13}$& $1549.14$ &  $1397.94$\\
		
			 NGC 3377 &$ 1.78\times 10^{8} $&$ 10.99$
&$7.726\times 10^{-13}$ & $593.005$ &  $535.26$\\
		
			NGC 4697 &$  2.02\times 10^{8}  $&$ 12.54  $ &$7.684\times 10^{-13}$& $672.96$ &  $607.28$\\
			 
			 NGC 5128 &$  5.69\times 10^{7}  $& $3.62   $ &$7.498\times 10^{-13}$& $189.562$ &  $171.06$\\
			
			NGC 1316&$  1.69\times 10^{8}  $&$20.95   $ &$3.848\times 10^{-13}$& $563.021 $ &  $508.07$\\
			
			 NGC 3607 &$ 1.37\times 10^{8} $&$ 22.65  $ &$2.885\times 10^{-13}$& $456.414 $ &  $411.866$\\
			
			NGC 4473 &$  0.90\times 10^{8}  $&$ 15.25  $ &$2.815\times 10^{-13}$& $299.834$ &  $270.57$\\
			
			 NGC 4459 &$ 6.96\times 10^{7} $&$ 16.01  $ &$2.073\times 10^{-13}$ & $231.871 $ &  $209.24$\\
		
			M32 &$ 2.45\times 10^6$ &$ 0.8057 $
&$1.450\times 10^{-13}$ & $8.16214 $ &  $7.364$    \\
			
			 NGC 4486A &$ 1.44\times 10^{7} $&$ 18.36  $ &$3.741\times 10^{-14}$ & $47.9734$ &  $43.2912$\\
			 
			NGC 4382 &$  1.30\times 10^{7}  $&$ 17.88 $  &$3.468\times 10^{-14}$& $43.3093$ &  $39.0824$\\
		
			CYGNUS A &$  2.66\times 10^{9}  $&$ 242.7 $  &$1.4174\times 10^{-15}$& $8861.76$ &  $7996.85$\\
		\end{tabular}
	\end{ruledtabular}
\caption{ Estimation of time delay for supermassive black holes at the center of nearby galaxies in the case  Kerr ($a=0.8$) and RQCBH ($a=0.8$ and $\alpha=0.4$). Mass ($M$) and distance ($D_{OL}$) are given in the units of solar mass and Mpc, respectively. Time Delays are expressed in minutes. \bf{The masses and the distances of the central black holes are taken from \cite{EventHorizonTelescope:2022xqj,EventHorizonTelescope:2019dse,McConnell:2012hz,Kormendy:2013dxa}}
	}\label{table3} 
\end{table*}

\section{Strong gravitational lensing by supermassive black holes,  M87* and Sgr A*}\label{Sec4}
In this section, we model the supermassive black holes Sgr A* and M87* as the RQCBH spacetime to numerically estimate the strong lensing observables, $\theta_{\infty}$, separation $s$, and relative magnification $r_{\text{mag}}$ for different values of parameter $\alpha$ and also compare the values with that of Kerr black holes for different values of spin parameters $a$. We depict our results in Fig.~\ref{plottheta}, Fig.~\ref{plotseperation}  and Fig.~\ref{plotrmag},  while Table~\ref{table1} shows the estimated values for lensing observables for various values of $a$, and $\alpha$ in comparison with Schwarzschild~($a=\alpha=0$) and Kerr black hole~($\alpha=0$). Our study of RQCBH as a lens indicates that the angular positions of images for Sgr A* and M87* align well with the angular shadow diameters measured by EHT for both Sgr A* and M87*. The results in Table \ref{table1} show that in the case of RQCBH, the angular positions of the images are smaller than their corresponding values in Kerr. Using Eq.~\ref{angpos} we have also compared the primary and secondary angular position with the position of the source ($\beta$) in Fig.~\ref{comparion}. In comparison, it shows that $|\theta_p|>|\theta_s|$, always, also the difference between primary and secondary images decreases as the order($n$) of the angular position increases. Furthermore, the angular separation $s$ increases with $\alpha$ at lower spin levels while it increases and then after a certain value decreases with $\alpha$ for higher spin levels (c.f. Fig.~\ref{plotseperation}). The separation $s$ in the case of RQCBH for Sgr A* and M87* range between 0.033-0.792 $\mu$as and 0.025-0.592$\mu$as, respectively. Table.~\ref{tab2} shows the deviation of the lensing observables of RQCBH from the Kerr black hole. The quantum-corrected parameter $\alpha$ shows a higher deviation for higher spin values with deviation in image position reaching as much as $\sim1.6~\mu$as and $\sim1.2~\mu$as for Sgr A* and M87*, respectively. The relative magnification of the first-order images of RQCBHs is less magnified than the corresponding images of black holes in GR. The magnification decreases slowly with the parameter $\alpha$ for lower spin levels and for high spin values $r_{mag}$ decrease rapidly with the parameter $\alpha$, as shown in Table \ref{table1} and Fig.~\ref{plotrmag}. Using Eq.~\ref{mag}, we see that the magnification is inversely proportional to the position of the source ($\beta$) (c.f. Fig.~\ref{comparion}). We have also calculated the time delay for different black holes in nearby galaxies in Table~\ref{table3}.  The time delay of the first image from that of the second image, $\Delta T_{2,1}$, for RQCBH as Sgr A* and M87 * can reach $24.95$~ min and $308.15$~ h, respectively, while the deviation from the Kerr black hole for Sgr A* and M87 * is $2.69$~ min and $33.327$~ h, respectively.  Observing the time delay in Sgr A* is much shorter and more difficult to measure, and we have to wait for the next-generation event horizon telescope (ngEHT) to do the same.

\section{Constraints from EHT observations of M87* and Sgr A*}\label{Sec5}
\begin{figure*}[tbh!]
	\begin{centering}
		\begin{tabular}{p{9cm} p{9cm}}
		    \includegraphics[scale=0.75]{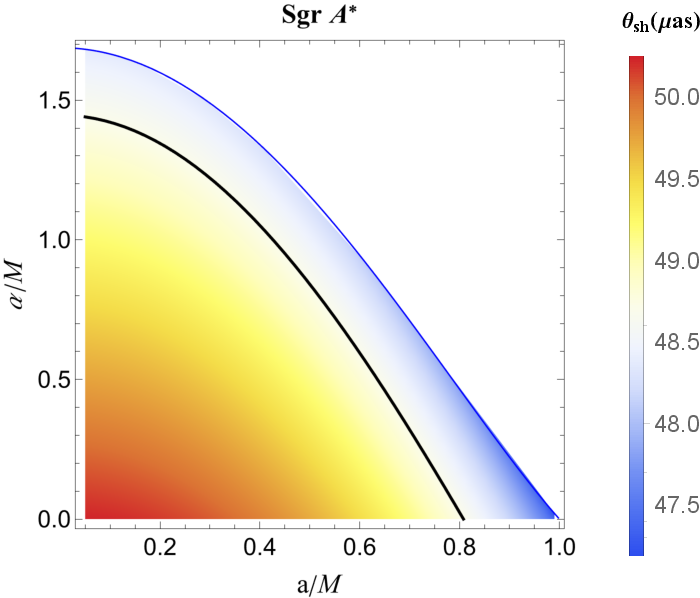}&
			\includegraphics[scale=0.75]{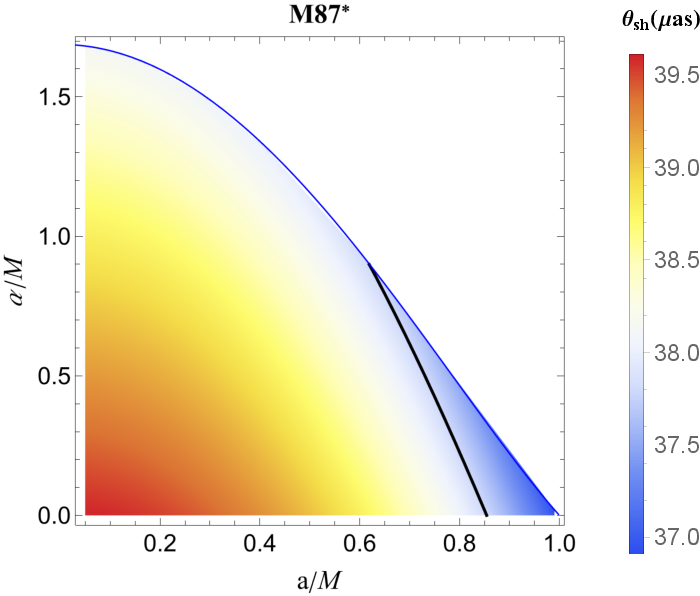}			
		\end{tabular}
	\end{centering}
	\caption{ {Shadow angular diameter $\theta_{sh}=2\theta_{\infty}$ of RQCBH as a function of $(a,\alpha)$. The solid black line correspond to the Sgr A* black hole shadow at $\theta_{sh}=48.7~\mu$as such that the region under this line satisfies the Sgr A* shadow 1-$\sigma$ bound (\textit{left}). M87* shadow angular diameter when considered as an RQCBH. The black line is $\theta_{sh}=37.8~\mu$as, and the region contained within it meets the M87* shadow 1-$\sigma$ bound (\textit{right}). The solid blue line in both figures corresponds to the parameter space.}}\label{EHT}		
\end{figure*}

Black hole shadows are the distinctive dark region encircled by a bright ring called photon rings \cite{bardeen1973black,cunningham1973optical}, which emerges from the interaction of photons with the intense gravitational field of the black hole. These photon rings are intricately linked to the black hole's specific parameters \cite{Johannsen:2010ru}. To unravel the implications of these shadows for near-horizon geometry, a considerable amount of analytical and numerical research has been dedicated to examining and modeling them \cite{Falcke:1999pj,Shen:2005cw,Yumoto:2012kz,Atamurotov:2013sca,Abdujabbarov:2015xqa,Cunha:2018acu,Kumar:2018ple,Afrin:2021ggx,Hioki:2009na, Amarilla:2010zq,Amarilla:2011fx,Amarilla:2013sj,Amir:2017slq,Singh:2017vfr,Mizuno:2018lxz,Allahyari:2019jqz,Papnoi:2014aaa,Kumar:2020hgm,Kumar:2020owy,Ghosh:2020spb,Afrin:2021wlj,Vagnozzi:2022moj,Vagnozzi:2019apd,Afrin:2021imp,Jusufi:2020cpn,Nampalliwar:2021tyz,Jusufi:2022loj,Jafarzade:2023dak}. 
The Event Horizon Telescope (EHT) observations have released the shadows images of the supermassive black holes Sgr A*\cite{EventHorizonTelescope:2019dse} and M87* \cite{EventHorizonTelescope:2022xqj}, with their measured sizes aligning with Kerr black hole forecasts within a 10\% accuracy. This remarkable accuracy provides a powerful tool for probing the nature of strong-field gravity.
 It allows for the imposition of constraints on possible deviations from the Kerr model, including effects like quantum effects in the strong-field regime. Thus, by considering the M87* and Sgr A* as RQCBH black holes, we can analyse how the EHT observation constrains the RQCBH-black hole parameters. By using the lensing observable ($\theta_{\infty}$) as the angular size of the black hole shadow, we confine the parameters ($a$, $\alpha$) within the 1-$\sigma$ region. 
\paragraph{Constraints from  M87* shadow}
In 2019, the Event Horizon Telescope (EHT) collaboration captured the first image of the supermassive black hole M87*, revealing a ring with a diameter of $\theta_{sh} = 42 \pm 3, \mu$as \cite{EventHorizonTelescope:2019dse}. Analysis shows that regardless of the parameters $a$ and $\alpha$, the Kerr black hole model gives its mass of $M = (6.5 \pm 0.7) \times 10^9 M_\odot$ and distance of $D_{OL} = 16.8$ Mpc and produces the shadow that fits within the 1-$\sigma$ confidence region of the observed data\cite{EventHorizonTelescope:2019dse,EventHorizonTelescope:2019pgp,EventHorizonTelescope:2019ggy}.  {By accounting for the offset between the emission ring diameter and angular shadow diameter of $\le$10\% \cite{EventHorizonTelescope:2019dse}, the mean angular shadow diameter comes out to be $\approx 37.8~\mu as$. Fig.~\ref{EHT} depicts the angular diameter $\theta_{sh}$ as a function of ($a,\alpha$) for the RQCBH as M87*, with the black corresponding to $\theta_{sh}=37.8~\mu$as. The 1-$\sigma$ bounds the RQCBH parameters ($a, \alpha$), viz., $0.615\le a \le 0.851$ and $0< \alpha \le 0.898$ and for $a<0.615$, there is no constraint on the parameter $\alpha$}. Thus, based on Fig.~\ref{EHT}, RQCBH can be a candidate for the astrophysical black holes.

\paragraph{Constraints from  Sgr A* shadow}
 {Before we proceed, we point out that the spin of Sgr A* has been widely debated. Early estimates suggested a lower spin \cite{Huang:2009sq, Broderick:2016ewk, Broderick:2010kx, Fragione:2020khu}, attributed to lower activity levels and interstellar scattering. More recent studies, however, suggest a higher spin around $a=0.9$ \cite{Daly:2023axh, Afrin:2023uzo}, indicating near-maximal rotation. EHT's 2022 observations found spin values around $a=0.5$ and $a = 0.94$ \cite{EventHorizonTelescope:2022xqj}, suggesting Sgr A*'s spin may range between $a=0.65$ and $a=0.9$ \cite{Dokuchaev:2023obv}.} 
The EHT's 2022 observations of Sgr A* estimated the shadow diameter to be $\theta_{sh} = (48.7 \pm 7) \mu$as and the emission ring's angular diameter to be $\theta_d = (51.8 \pm 2.3) , \mu$as, using previous estimates of the black hole's mass as $M = 4.0^{+1.1}{-0.6} \times 10^6 M\odot$ and distance as $D_{LS} = 8.15 \pm 0.15$ kpc \cite{EventHorizonTelescope:2022xqj}. Utilizing three independent algorithms, EHT-imaging, SIMLI, and DIFMAP, the shadow diameter was averaged between $\theta_{sh} = (46.9, 50) \mu$as.  {The mean angular shadow value is 48.7$\mu$as which strongly constrains the parameters $a$ and $\alpha$ for the RQCBH, which falls within the $1- \sigma$ confidence region with the observed angular diameter such that $0< a \le 0.806$ and $0< \alpha \le 1.443$ are allowed for the RQCBH.} Thus, within the finite parameter space, RQCBH agrees with the EHT results of Sgr A* black hole shadow (cf. Fig.~\ref{EHT}).  

 { Thus, our analysis aligns with a higher spin estimate of $a \approx 0.8$. Determining the precise spin vector remains challenging due to limitations in telescope coverage, systematics, and accretion modelling. With new data from ngEHT \cite{galaxies11010002}, future measurements should allow a more accurate determination of Sgr A*'s spin and related parameters.}

\section{Conclusions}\label{Sec6}
A mathematically compatible model of rotating black holes, including quantum gravity effects, remains inaccessible, posing challenges for testing quantum gravity predictions against observational data, such as those from the Event Horizon Telescope (EHT). The EHT's observations of Sgr A* and M87* are consistent with Kerr black holes as predicted by general relativity. Still, the growing number of observations of rotating black holes highlights the need for quantum-corrected models.
We derived a rotating quantum-corrected black hole (RQCBH) solution by applying a modified Newman-Janis algorithm (NJA) to a quantum-corrected non-rotating seed metric (\ref{1}). The resultant RQCBH retains a Kerr-like structure, with an additional quantum correction parameter $\alpha$ apart from mass $M$ and spin $a$. This solution allows for exploring quantum gravitational effects in the strong-field regime and allows constraints on $\alpha$ using EHT data.
We focus on gravitational lensing in the strong-deflection limit to assess how the quantum correction parameter $\alpha$ affects key observables, including the deflection angle $\alpha_D(\theta)$, angular position $\theta_{\infty}$, angular separation $s$, magnification ratio $r_{\text{mag}}$, and critical impact parameter $u_m$.
 
Our analysis reveals that the unstable photon orbit radius $x_{ps}$, the critical impact parameter $u_{ps}$ decreases with $\alpha$. We also found a decrease in the deflection angle with $\alpha$ such that RQCBH leads to a smaller deflection angle than the Kerr black hole, and the deflection angle diverges at smaller $u_{ps}$ for larger values of both $\alpha$ and $a$.
 
By considering gravitational lensing around supermassive black holes such as Sgr A* and M87* within the framework of the RQCBH spacetime, we observed significant deviations from general relativity (GR) predictions. In particular, we observe a decrease in the angular radius of relativistic images ($\theta_{\infty}$) with the parameter $\alpha$. For Sgr A*, $\theta_{\infty}$ ranges between 14.803 and 26.33 $\mu as$, with deviations from GR predictions reaching up to 1.6 $\mu as$. In contrast, for M87*, $\theta_{\infty}$ ranges from 11.12 to 19.782 $\mu as$, with deviations up to 1.2 $\mu as$ for a spin value of $a=0.8$. The angular separation $s$ increases rapidly with $\alpha$ at lower spin levels, while after a certain value, it also starts to decrease with $\alpha$ for higher spin levels. The separation $s$ in the case of RQCBH for Sgr A* and M87* range between 0.033-0.792 $\mu$as and 0.025-0.592$\mu$as, respectively, and deviations from the Kerr black hole can be as large as 0.416 $\mu as$ for Sgr A* and 0.31 $\mu as$ for M87* for spin value of $a=0.8$.

Additionally, the magnification of the first-order images is lower in the RQCBH spacetime compared to Kerr black holes. The relative magnification decreases with $\alpha$ with relative magnitudes $r_{mag}$ ranging from 1.817 to 6.822. Time delays between the second and first images, denoted as $\Delta T_{2,1}$, exhibit substantial deviations from Kerr black hole predictions, reaching up to  {6.0127 min} for Sgr A* and 308.15 h for M87*.  {The Event Horizon Telescope (EHT) measurements constrain the RQCBH parameters, showing that for M87*, when the spin parameter $a<0.615$, there is no constrain on the value of parameter $\alpha$, for $a\in(0.615,0.851)$, the $\alpha$ go up to 0.898, while for Sgr A*, parameters are constrained as $0< a \le 0.806$ and $0< \alpha \le 1.443$}. These findings impose stricter constraints on the RQCBH parameters for Sgr A* than for M87*. Our findings also reveal that EHT data constrain the SgrA* to have a higher spin value. 

The results presented here generalize previous discussions on black hole lensing in GR. Our results go over to Kerr and Schwarzschild black holes in the limits $\alpha \to 0$ and $a,\alpha \to 0$, respectively. Although resolving the order estimated in the SDL is challenging, the outlook for future observations looks bright.

\section{Acknowledgments} 
A.V and S.G.G. would like to thank SERB-DST for project No. CRG/2021/005771.

\section{Data Availability Statements}
No new data were generated or analysed in support of this research, nor was any third-party data analyzed.
\bibliography{citation}
\end{document}